\documentclass[a4paper,11pt]{article}
\pdfoutput=1% if your are submitting a pdflatex (i.e. if you have
\usepackage{jheppub} % for details on the use of the package, please
\usepackage{graphicx}
\usepackage{url}
\usepackage{bigdelim}
\usepackage{booktabs}
\usepackage{dcolumn}
\usepackage{multirow}
\usepackage[large]{subfigure}
\usepackage{cancel}
\usepackage{stackrel}
\usepackage{paralist}
\usepackage{xspace}
\usepackage{slashed}
\usepackage[T1]{fontenc} % if needed
\usepackage[sc]{mathpazo}
\newcommand{\nua}[1]{\ensuremath{\rlap{\kern-2.5pt\ensuremath{\overset{\scriptscriptstyle(-)}{\phantom{\nu}}}}{\ensuremath{{\nu}_{#1}}}}}
\newcommand{\vet}[1]{\ensuremath{\hskip-1pt\vec{\hskip1pt#1}}}
\usepackage{enumitem}

\newcommand{\cenns}{CE$\nu$NS\xspace}

\begin{document}

\title{Constraints on light vector mediators through coherent elastic neutrino nucleus scattering data from COHERENT}

\author[a]{M. Cadeddu,}
\author[b]{and N. Cargioli,}
\author[a]{F. Dordei,}
\author[c]{C. Giunti,}
\author[d,e]{Y.F. Li,}
\author[a,b]{E. Picciau,}
\author[d,e]{and Y.Y. Zhang}

% The "\note" macro will give a warning: "Ignoring empty anchor..."
% you can safely ignore it.

\affiliation[a]{Istituto Nazionale di Fisica Nucleare (INFN), Sezione di Cagliari,
Complesso Universitario di Monserrato - S.P. per Sestu Km 0.700,
09042 Monserrato (Cagliari), Italy}
\affiliation[b]{Dipartimento di Fisica, Universit\`{a} degli Studi di Cagliari,
Complesso Universitario di Monserrato - S.P. per Sestu Km 0.700,
09042 Monserrato (Cagliari), Italy}
\affiliation[c]{Istituto Nazionale di Fisica Nucleare (INFN), Sezione di Torino, Via P. Giuria 1, I--10125 Torino, Italy}
\affiliation[d]{Institute of High Energy Physics, Chinese Academy of Sciences, Beijing 100049, China}
\affiliation[e]{School of Physical Sciences, University of Chinese Academy of Sciences, Beijing 100049, China}

% e-mail addresses: one for each author, in the same order as the authors
\emailAdd{matteo.cadeddu@ca.infn.it}
\emailAdd{nicolacargioli@gmail.com}
\emailAdd{francesca.dordei@cern.ch}
\emailAdd{carlo.giunti@to.infn.it}
\emailAdd{liyufeng@ihep.ac.cn}
\emailAdd{emmanuele.picciau@ca.infn.it}
\emailAdd{zhangyiyu@ihep.ac.cn}

%\date{\dayofweekname{\day}{\month}{\year} \ddmmyydate\today, \currenttime}
%\date{5 November 2019}

\abstract{We present new constraints on three different models, the so-called universal, $B-L$ and $L_\mu-L_\tau$ models, involving a yet to be observed light vector $Z'$ mediator, by exploiting the recent observation of coherent elastic neutrino-nucleus scattering (\cenns) in argon and cesium-iodide performed by the COHERENT Collaboration. 
We compare the results obtained from a combination of the above data sets with the limits derived from searches in fixed target, accelerator, solar neutrino and reactor \cenns experiments, and with the parameter region that could explain the anomalous magnetic moment of the muon.
We show that for the universal and the $B-L$ models, the COHERENT data allow us to put stringent limits in the light vector mediator mass, $M_{Z'}$, and coupling, $g_{Z'}$, parameter space.}

\maketitle
\flushbottom

\section{Introduction}
\label{sec:introduction}

Since the first observation of the coherent elastic neutrino-nucleus scattering (\cenns) process in cesium-iodide (CsI) by the COHERENT experiment~\cite{Akimov:2017ade,Akimov:2018vzs}, many intriguing physics results have been derived by a large community of physicists~\cite{Cadeddu:2017etk,Papoulias:2019lfi,Coloma:2017ncl,Liao:2017uzy,Kosmas:2017tsq,Denton:2018xmq,AristizabalSierra:2018eqm,Cadeddu:2018dux,Dutta:2019eml,Dutta:2019nbn,papoulias2019coherent,Khan_2019,Cadeddu_2019,Cadeddu:2019eta} on very diverse physical sectors. With the recent detection of \cenns in a single-phase 24 kg liquid-argon scintillation detector~\cite{Akimov:2020pdx} the COHERENT Collaboration has started to probe the \cenns cross-section dependence on the square of the number of neutrons, $N^2$, and unrevealed a new way to test the standard model (SM). Indeed, this new measurement allowed to gain additional and complementary information to that provided by the CsI dataset on nuclear physics, neutrino properties, physics beyond the SM, and electroweak interactions~\cite{Cadeddu:2020lky}.

\cenns is a neutral current process induced by the exchange of a $Z$ boson. It thus represents also a sensitive probe for non standard interactions (NSI) that are not included in the SM~\cite{Giunti:2019xpr}, induced by yet to be discovered neutral vector bosons~\cite{Liao:2017uzy}, particularly if they are light.
Indeed, for sufficiently light vector mediator masses,  the  scattering  rate  grows as $1/|\vec{q}|^2$ as one goes to lower energies, so the low momentum transfer of \cenns experiments makes them ideal laboratories for such searches.
In fact, \cenns can occur when the three-momentum transfer $|\vec{q}| \simeq \sqrt{2 M T_\mathrm{nr}}$ during neutrino scattering off a nucleus is smaller than the inverse of the nuclear radius $R$, which is of the order of few fm, where $M$ is the nuclear mass, and $T_\mathrm{nr}$ is the energy of the nuclear recoil of a few keV. This means that on average the momentum transfer is of the order of few tens of MeV, making \cenns the perfect place to study scenarios including a new light vector mediator. The nature of the latter depends on the details of the specific model assumed.

%The observation in argon, which is the lightest nucleus for which \cenns process has been measured,
%In the process, the scattering amplitudes of the nucleons inside a nucleus are in phase and addcoherently, which leads to a large enhancement of the cross section.  In the SM, CE谓NS is inducedvia the exchange of aZboson [10].  Hence,  CE谓NS is also sensitive to NSI induced by a newneutral vector boson [11, 12].  To probe NSI at higher energies, the formulation of Eq. (1.1) mayno longer be valid.  First, the momentum-dependent propagator of the mediator should be usedif its massMis not much larger than the typical momentum transferQto properly model theenergy dependence.  Second, SM gauge-invariant operators must be adopted when the momentumtransfer or the mediator mass is at or above the electroweak scale.

In this paper, we present new constraints on different models involving
a light vector $Z'$ mediator, obtained analyzing the
new COHERENT argon (Ar) and CsI data, as well as those obtained with a combined analysis of these two datasets,
using for CsI the same inputs described in Ref.~\cite{Cadeddu:2019eta}. In particular, we consider three models with different interactions of the
light vector $Z'$ mediator. The first one is the so-called universal model, in which the mediator couples universally to all the SM fermions~\cite{Liao:2017uzy,Billard_2018}. The second one is referred to as the $B-L$ model~\cite{Liao:2019uzy,Billard_2018}, where the coupling of the mediator is different between quarks and leptons. Finally, the third one is the so-called $L_\mu-L_\tau$ model~\cite{Altmannshofer:2019zhy}, in which the mediator only couples with SM particles of the muonic or tauonic flavour. All these models are theoretically well motivated to provide a coherent explanation to a series of emerging discrepancies in precision studies of low-energy observables. Among those, the anomalous measurement of the magnetic moment of the muon, referred to as $(g-2)_\mu$, performed by the E821 experiment at BNL~\cite{Bennett_2006} represents since almost two-decades an intriguing puzzle. Indeed, the experimental value differs from the SM prediction by about~3.7$\sigma$~\cite{aoyama2020anomalous}. Thus, new constraints on this kind of models that incorporate new physics in the leptonic sector are very much awaited. In this context, the new COHERENT results provide a timely and stimulating opportunity to probe some of these models and improve the existing limits.

The plan of the paper is as follows.
In Section~\ref{sec:theory} we describe briefly the \cenns formalism used to simulate the \cenns signal in the SM as well as the experimental inputs.
In Section~\ref{sec:light_med} we summarise the models of interest to incorporate the effect of a light vector $Z'$ mediator on the \cenns cross section while in Section~\ref{sec:data_analysis} we describe our method to analyse the COHERENT data. In Section~\ref{sec:const} we derive the constraints using the results of the analysis of the COHERENT Ar and CsI data as well as the combined limits and we compare them with the existing limits.
Finally, in Section~\ref{sec:conclusions} we summarise the results of the paper.

\section{\cenns signal prediction in the standard model}
\label{sec:theory}

We follow closely the same formalism developed in Ref.~\cite{Cadeddu:2020lky} to analyse \cenns Ar data and in Ref.~\cite{Cadeddu:2019eta} for CsI data from the COHERENT Collaboration, to which the reader is referred for the details.
The SM weak-interaction differential cross section as a function of the nuclear kinetic recoil energy $T_\mathrm{nr}$
of \cenns processes with a spin-zero nucleus $\mathcal{N}$ with $Z$ protons and $N$ neutrons is given by~\cite{Drukier:1983gj,Barranco:2005yy,Patton:2012jr}
\begin{equation}
\dfrac{d\sigma_{\nu_{\ell}\text{-}\mathcal{N}}}{d T_\mathrm{nr}}
(E,T_\mathrm{nr})
=
\dfrac{G_{\text{F}}^2 M}{\pi}
\left(
1 - \dfrac{M T_\mathrm{nr}}{2 E^2}
\right)
Q_{\ell,\,\mathrm{SM}}^2
%\left[
%g_{V}^{p}(\nu_{\ell})
%Z
%F_{Z}(|\vet{q}|^2)
%+
%g_{V}^{n}
%N
%F_{N}(|\vet{q}|^2)
%\right]^2
,
\label{cs-std}
\end{equation}
where $G_{\text{F}}$ is the Fermi constant, $\ell = e, \mu, \tau$ denotes the neutrino flavour, $E$ is the neutrino energy and $Q_{\ell,\,\mathrm{SM}}^2=\left[
g_{V}^{p}(\nu_{\ell})
Z
F_{Z}(|\vet{q}|^2)
+
g_{V}^{n}
N
F_{N}(|\vet{q}|^2)
\right]^2$.
For the neutrino-proton, $g_{V}^{p}$, and the neutrino-neutron, $g_{V}^{n}$, couplings we consider the more accurate values that take into account
radiative corrections in the minimal subtraction, $\overline{\text{MS}}$, scheme~\cite{Erler:2013xha,Cadeddu:2020lky}, that correspond to
\begin{align}
\null & \null
g_{V}^{p}(\nu_{e})
=0.0401
,
\label{gVp-nue}
\\
\null & \null
g_{V}^{p}(\nu_{\mu})
=0.0318
,
\label{gVp-num}
\\
\null & \null
g_{V}^{n}
=-0.5094
.
\label{gVn}
\end{align}
These values are different from the tree-level values
$g_{V}^{p}=0.0229$
and
$g_{V}^{n}=-0.5$,
in particular for those of $g_{V}^{p}(\nu_{e})$ and $g_{V}^{p}(\nu_{\mu})$.

In Eq.~(\ref{cs-std})
$F_{Z}(|\vet{q}|^2)$
and
$F_{N}(|\vet{q}|^2)$
are, respectively, the form factors of the proton and neutron distributions in the nucleus.
They are given by the Fourier transform of the corresponding nucleon
distribution in the nucleus and
describe the loss of coherence for
$|\vet{q}| R_{p} \gtrsim 1$
and
$|\vet{q}| R_{n} \gtrsim 1$,
where $R_{p}$ and $R_{n}$ are, respectively, the rms radii of the proton and neutron distributions.
For the form factors of the proton and neutron distributions we employ the
Helm parameterisation~\cite{Helm:1956zz},
that is practically equivalent to the other two commonly-used
symmetrized Fermi~\cite{Piekarewicz:2016vbn}
and
Klein-Nystrand~\cite{Klein:1999qj}
parameterisations.
The description of these parameterisations can be found in
several papers,
for example in Refs.~\cite{Piekarewicz:2016vbn,Cadeddu:2017etk,Khan_2019,papoulias2019coherent,Cadeddu:2020lky}.
For the proton rms radii, in our calculations we use~\cite{fricke,Angeli:2013epw}
\begin{align}\nonumber
R_{p}\,(\mathrm{Cs}) = 4.804&\, \text{fm},\quad
R_{p}\,(\mathrm{I}) = 4.749  \, \text{fm},\\
R_{p}\,(\mathrm{Ar}) &= 3.448 \, \text{fm}.
\label{Rp}
\end{align}
The value of the neutron rms radius is only poorly known experimentally both in CsI~\cite{Cadeddu:2017etk,Papoulias:2019lfi,Cadeddu:2018dux,Huang:2019ene,papoulias2019coherent,Khan_2019,Cadeddu:2019eta} and Ar~\cite{Cadeddu:2020lky}. For CsI, we adopt the values $R_{n}\,(\mathrm{Cs}) = 5.01 \, \text{fm}$ and  $R_{n}\,(\mathrm{I}) = 4.94 \, \text{fm}$ obtained with the relativistic mean field (RMF) NL-Z2~\cite{PhysRevC.60.034304} nuclear model calculation in Ref.~\cite{Cadeddu:2017etk}, that are in good agreement with the most precise experimental value determined in combination with atomic parity violation (APV) experimental results in Ref.~\cite{Cadeddu:2019eta}. For argon, it is obtained starting from the experimental value of the proton rms radius and considering the average value of the so-called neutron skin, that corresponds to the difference among the two radii, predicted by different models with diverse nuclear interactions~\cite{Reinhard:1995zz,Chabanat:1997un,Kim-Otsuka-Bonche-1997,Klupfel:2008af,Kortelainen:2010hv,Kortelainen:2011ft,Bartel:1982ed,Dobaczewski:1983zc,Niksic:2008vp,Niksic:2002yp,PhysRevC.100.061304}. The neutron skin is predicted to be between 0.06 and 0.11 fm. Considering a neutron skin of 0.1~fm that is predicted by most of the models, we get $R_{n}\,(\mathrm{Ar}) = 3.55 \, \text{fm}$.

The CE$\nu$NS event rate in the COHERENT experiment~\cite{Akimov:2017ade,Akimov:2020pdx}
depends on the neutrino flux produced from the Spallation Neutron Source (SNS) at the Oak Ridge National Laboratories.
The total differential neutrino flux is given by the sum of the three neutrino components coming from the pion decay ($\pi^+\to \mu^++\nu_\mu$) and the subsequent muon decay ($\mu^{+} \to e^{+} + \nu_{e} + \bar\nu_{\mu}$)
\begin{align}
\frac{d N_{\nu_{\mu}}}{d E}
=
\null & \null
\eta
\,
\delta\!\left(
E - \dfrac{ m_{\pi}^2 - m_{\mu}^2 }{ 2 m_{\pi} }
\right)
,
\label{numu}
\\
\frac{d N_{\nu_{\bar\mu}}}{d E}
=
\null & \null
\eta
\,
\dfrac{ 64 E^2 }{ m_{\mu}^3 }
\left(
\dfrac{3}{4} - \dfrac{E}{m_{\mu}}
\right)
,
\label{numubar}
\\
\frac{d N_{\nu_{e}}}{d E}
=
\null & \null
\eta
\,
\dfrac{ 192 E^2 }{ m_{\mu}^3 }
\left(
\dfrac{1}{2} - \dfrac{E}{m_{\mu}}
\right).
\label{nue}
\end{align}
Here $m_\pi$ and $m_\mu$ are the pion and muon masses, and the normalization factor is
$\eta = r N_{\text{POT}} / 4 \pi L^2 $,
where $r$ is the number of neutrinos per flavour
that are produced for each proton-on-target (POT), $ N_{\text{POT}}$ is the number of proton on target, and $L$ is the distance between the source and the detector.
For the dataset collected by the COHERENT Ar detector, called CENNS-10~\cite{Akimov:2017ade}, we use $r=(9\pm0.9)\times10^{-2}$,
$ N_{\text{POT}} = 13.7\times10^{22} $
that corresponds to a total integrated beam power of 6.12 GW$\cdot$hr
and $L = 27.5 \, \text{m}$.
The same numbers for CsI are $r=0.08$, $ N_{\text{POT}} = 17.6\times10^{22} $ and $L = 19.3 \, \text{m} $.
The pions decay at rest producing $\nu_\mu$'s which arrive at the COHERENT detector as a prompt signal within about
$1.5 \, \mu\text{s}$
after protons-on-target. The decay at rest of $\mu^{+}$ produces
a delayed component of $\bar\nu_{\mu}$'s and $\nu_e$'s, since they arrive at the detector in a relatively longer time interval of about $10 \, \mu\text{s}$.
The theoretical \cenns event-number $N_i^{\mathrm{CE}\nu\mathrm{NS}}$ in each nuclear-recoil energy-bin $i$ is given by\footnote{Note that for CsI the two contributions sum incoherently, thus $ \dfrac{d\sigma_{\nu\text{-}\mathrm{CsI}}}{d T_\mathrm{nr}}(E,T_\mathrm{nr})=\dfrac{d\sigma_{\nu\text{-}\mathrm{Cs}}}{d T_\mathrm{nr}}(E,T_\mathrm{nr})+\dfrac{d\sigma_{\nu\text{-}\mathrm{I}}}{d T_\mathrm{nr}}(E,T_\mathrm{nr})$}
\begin{equation}
\label{eq:en_spectra}
 N_i^{\mathrm{CE}\nu\mathrm{NS}} = N(\mathcal{N}) \int^{T_{\mathrm{nr}}^{i+1}}_{T_{\mathrm{nr}}^{i}} dT_{\mathrm{nr}}\,A(T_{\mathrm{nr}}) \int^{E_{\mathrm{max}}}_{E_{\mathrm{min}}} dE \sum_{\nu=\nu_e,\nu_\mu,\overline{\nu}_\mu} \frac{dN_\nu}{dE}  \dfrac{d\sigma_{\nu\text{-}\mathcal{N}}}{d T_\mathrm{nr}}(E,T_\mathrm{nr})\,,
\end{equation}
where $A(T_{\mathrm{nr}})$ is the energy-dependent reconstruction efficiency, $E_{\mathrm{min}} = \sqrt{MT_\mathrm{nr}/2}$, $E_{\mathrm{max}} = m_\mu/2 \sim 52.8$ MeV, and $dN_\nu/dE$ is the neutrino flux integrated over the experiment lifetime. The electron-equivalent recoil energy $T_{ee}\,[\mathrm{keV}_{ee}]$, is transformed into the nuclear recoil energy $T_{\mathrm{nr}}\,[\mathrm{keV}_{\mathrm{nr}}]$ thanks to the relation
\begin{equation}
\label{eq:qf}
T_{ee} = f_Q (T_{\mathrm{nr}})T_{\mathrm{nr}}\,,
\end{equation}
where $f_Q$ is the quenching factor, which is the ratio between the scintillation light emitted in nuclear and electron recoils.
%and determines the relation between the number of detected photoelectrons and the nuclear recoil kinetic energy.
In CsI, the quenching factor is taken from Fig. 1 in Ref.~\cite{Collar:2019ihs}. This represents the most precise determination of the quenching factor in CsI up to date.
%, however for completeness in Appendix~\ref{app:qfcoherent} we show the results obtained also with the official, but significantly less precise, quenching factor released by the COHERENT collaboration.
For Ar, following Ref.~\cite{Akimov:2020pdx}, the quenching factor is parameterised as $f_Q (T_{\mathrm{nr}}) = (0.246\pm0.006\, \mathrm{keV}_{\mathrm{nr}}) + ((7.8\pm0.9)\times10^{-4})T_{\mathrm{nr}}$ up to $125\,\mathrm{keV}_{\mathrm{nr}}$, and kept constant for larger values. Finally, the value of $N(\mathrm{CsI})$ is given by $N_\mathrm{A}\,M_{\mathrm{det}}/M_{\mathrm{CsI}}$, where $N_\mathrm{A}$ is the Avogadro number, $M_{\mathrm{det}}$ is the detector active mass equal to 14.6 kg and $M_{\mathrm{CsI}}=259.8$ g/mol is the molar mass of CsI. Similarly, the value of $N(\mathrm{Ar})$ is obtained using $M_{\mathrm{det}}=24$~kg and $M_{\mathrm{Ar}}=39.96$~g/mol for $^{40}\mathrm{Ar}$. Here, the contribution of $^{36}\mathrm{Ar}$ and $^{38}\mathrm{Ar}$ has been neglected.

\section{Light vector $Z'$ mediator in \cenns}
\label{sec:light_med}

The SM cross section presented in Section~\ref{sec:theory} is modified by the presence of a new massive vector mediator which couples to SM leptons and quarks. Considering a vector neutral-current neutrino non-standard interaction~\cite{Dev:2019anc} and assuming that the neutrino does not change flavour, it
is generically described by the effective four-fermion
interaction Lagrangian
(see Ref.~\cite{Giunti:2019xpr} and references therein)
%\footnote{%
%\bluechange{%
%In general,
%neutral-current neutrino non-standard interactions can have also
%axial components,
%but their effect is negligible in coherent elastic scattering of neutrinos with heavy
%nuclei~\cite{Barranco:2005yy}.
%}
%}
\begin{equation}
\mathcal{L}_{\text{NSI}}^{\text{NC}}
=
- 2 \sqrt{2} G_{\text{F}}
\sum_{\ell=e,\mu}
\left( \overline{\nu_{\ell L}} \gamma^{\rho} \nu_{\ell L} \right)
\sum_{f=u,d}
\varepsilon_{\ell\ell}^{fV}
\left( \overline{f} \gamma_{\rho} f \right)
.
\label{lagrangian}
\end{equation}
The parameters
$\varepsilon_{\ell\ell}^{fV}$, where $f=u,d$ stands for the flavour of the quark and $\ell=e,\mu$ is the neutrino flavour\footnote{We consider only the first generation of quarks since they are the only ones contained in nuclei and only electronic and muonic neutrinos since they are the only species present in the flux at the SNS.}, describe the size of non-standard interactions relative to standard neutral-current weak interactions.
The full cross section comes from coherently summing the contributions from the exchange of the SM and NSI mediators, which may interfere.  In particular, the NSI mediator effectively induces an energy-dependent modification of the SM factor $Q_{\ell,\,\mathrm{SM}}^2
%=\left[
%g_{V}^{p}(\nu_{\ell})
%Z
%F_{Z}(|\vet{q}|^2)
%+
%g_{V}^{n}
%N
%F_{N}(|\vet{q}|^2)
%\right]^2$
$ in Eq.~\ref{cs-std}.
%To keep things as simple as possible, we assume that there is no $Z-Z鈥? mixing and that the $Z鈥? has a purely vector interaction with the fermions of the SM,
Indeed, considering the NSI scenario described before it becomes
\begin{align}
Q_{\ell}^2
=
\null & \null
\left[
\left( g_{V}^{p}(\nu_{\ell}) + 2 \varepsilon_{\ell\ell}^{uV} + \varepsilon_{\ell\ell}^{dV} \right)
Z
F_{Z}(|\vet{q}|^2)
+
\left( g_{V}^{n} + \varepsilon_{\ell\ell}^{uV} + 2 \varepsilon_{\ell\ell}^{dV} \right)
N
F_{N}(|\vet{q}|^2)
\right]^2.
%\nonumber
%\\
%\null & \null
%+
%\sum_{\beta\neq\alpha}
%\left|
%\left( 2 \varepsilon_{\alpha\beta}^{uV} + \varepsilon_{\alpha\beta}^{dV} \right)
%Z
%F_{Z}(|\vet{q}|^2)
%+
%\left( \varepsilon_{\alpha\beta}^{uV} + 2 \varepsilon_{\alpha\beta}^{dV} \right)
%N
%F_{N}(|\vet{q}|^2)
%\right|^2.
\label{Qalpha2}
\end{align}

In this paper, we focus on three simple models in which the vector neutral-current neutrino NSI is induced by a gauge boson $Z'$ with mass $M_{Z'}$ and coupling $g_{Z'}$ associated with a new $U(1)'$ symmetry.
In this scenario, the \cenns cross section can be determined by writing the parameter  $\varepsilon_{\ell\ell}^{fV}$ in terms of the light $Z'$ propagator as
\begin{equation}
    \epsilon_{\ell \ell}^{fV}=\dfrac{g_{Z'}^2\,Q'_\ell Q'_f}{\sqrt{2}G_F\, (|\vec{q}|^2+M_{Z'}^2)},
\end{equation}
where $Q'$ are the charges under the new gauge symmetry.

The first model that we consider describes a $Z'$ boson which couples universally to all SM fermions~\cite{Liao:2017uzy,Billard_2018}.
We set the charges to be $Q'_\ell \equiv Q'_f = 1$, and the coupling becomes the same for all the fermions.
Under this model, the cross section in Eq.~\ref{cs-std} becomes~\cite{Liao:2017uzy}
\begin{align}
    \Big(\dfrac{d\sigma}{dT_\mathrm{nr}}\Big)^{\nu_\ell\text{-}\mathcal{N}}_{\mathrm{univ}}(E,T_\mathrm{nr}) =
    \dfrac{G_F^2 M}{\pi} \Big(1-\dfrac{MT_\mathrm{nr}}{2E^2} \Big)
    \cdot
    \Big[Q_{\ell,\,\mathrm{SM}}+\dfrac{3(g^{}_{Z'})^2}{\sqrt{2}G_F}\dfrac{ZF_Z(|\vet{q}|^2)+NF_N(|\vet{q}|^2)}{|\vec{q}|^2+M_{Z'}^2}\Big]^2.
    \label{eq:uni}%\rm uni
\end{align}

The second model that we consider is the so-called $B-L$ (baryon number minus lepton number) extension of the SM~\cite{Liao:2019uzy,Billard_2018}. In this case the gauge charges are determined by imposing that the theory is anomaly free. In particular, in this model the boson couples universally to the quarks, as well as to the neutrinos, but with different charges, namely $Q'_\ell \neq Q'_f$. In particular, in the $B-L$ model the gauge charges are such that $Q'_\ell=1$ and $Q'_f=-Q'_\ell/3$.\\
Under these assumptions, the cross section in Eq.~\ref{cs-std} becomes
\begin{align}
    \Big(\dfrac{d\sigma}{dT_\mathrm{nr}}\Big)^{\nu_\ell\text{-}\mathcal{N}}_{\mathrm{B-L}}(E,T_\mathrm{nr})
    =\dfrac{G_F^2 M}{\pi} \Big(1-\dfrac{MT_\mathrm{nr}}{2E^2} \Big)
    \cdot
    \Big[Q_{\ell,\,\mathrm{SM}}-\dfrac{\left(g^{}_{Z'}\right)^2}{\sqrt{2}G_F}\dfrac{ZF_Z(|\vet{q}|^2)+NF_N(|\vet{q}|^2)}{|\vec{q}|^2+M_{Z'}^2}\Big]^2.
\end{align}%B-L

The last scenario that we consider is a model with gauged $L_\mu-L_\tau$ symmetry~\cite{Altmannshofer:2019zhy}. In this case, the new $Z'$ boson can couple directly only to muonic or tauonic flavour and there is no tree-level coupling to the quark sector.
Thus, this model can be studied through the \cenns process by considering the interaction between the new boson and quarks via kinetic loops of muons and tauons involving photons.
Under these assumptions, the cross section for this process becomes~\cite{Altmannshofer:2019zhy}
\begin{align}\nonumber
    &\Big(\dfrac{d\sigma}{dT_\mathrm{nr}}\Big)^{\nu_\ell\text{-}\mathcal{N}}_{L_\mu-L_\tau}(E,T_\mathrm{nr})
    =\dfrac{G_F^2 M}{\pi} \Big(1-\dfrac{MT_\mathrm{nr}}{2E^2} \Big)
    \cdot\\
    &\left\{ \left[ g_V^p(\nu_{\ell})-\dfrac{\alpha_{\rm EM}\, \left(g^{}_{Z'}\right)^2}{3\sqrt{2}\pi G_F} \log{\Big(\dfrac{m_\tau^2}{m_\mu^2} \Big)\dfrac{1}{|\vec{q}|^2+M_{Z'}^2}}\right] ZF_Z(|\vet{q}|^2)+g_V^n N F_N(|\vet{q}|^2) \right\}^2,
\end{align} %L_\mu-L_\tau
where $\alpha_{\rm EM}$ is the electromagnetic fine-structure constant
and $m_\tau$ is the tau lepton mass.
As visible, only protons interact with the new boson due to the presence of the photon in the loop, and only the proton coupling is modified with a term that is proportional to the electric charge. As far as the \cenns process in COHERENT is concerned, this last model  modifies only the muonic neutrino and antineutrino cross sections, leaving the electronic neutrino cross section unchanged, since there is no direct coupling to the electronic flavour.
It is worth to specify that in the case of antineutrinos, the term induced by the introduction of the light vector mediator changes sign as well as the other vector couplings.

\section{COHERENT data analysis}
\label{sec:data_analysis}

Concerning the analysis of the CsI COHERENT dataset, we considered the least-squares function
\begin{eqnarray}
\chi^2_{\text{CsI}}
&=&
\sum_{i=4}^{15}
\left(
\dfrac{
N_{i}^{\text{exp}}
-
\left(1+\alpha_{\text{c}}\right) N_{i}^{\mathrm{CE}\nu \mathrm{NS}}
-
\left(1+\beta_{\text{c}}\right) B_{i}
}{ \sigma_{i} }
\right)^2\\ \nonumber
&+&
\left( \dfrac{\alpha_{\text{c}}}{\sigma_{\alpha_{\text{c}}}} \right)^2
+
\left( \dfrac{\beta_{\text{c}}}{\sigma_{\beta_{\text{c}}}} \right)^2
+
\left( \dfrac{\eta-1}{\sigma_{\eta}} \right)^2
.
\label{chi-CsI}
\end{eqnarray}
For each energy bin $i$,
$N_{i}^{\text{exp}}$ is the experimental event number in Ref.~\cite{Akimov:2018vzs},
$N_{i}^{\mathrm{CE}\nu \mathrm{NS}}$
is the theoretical event number
that is calculated as explained in Sections~\ref{sec:theory},
$B_{i}$ is the estimated number of background events, and
$\sigma_{i}$ is the statistical uncertainty, both taken from Ref.~\cite{Akimov:2018vzs}.
As explained in Ref.~\cite{Cadeddu:2019eta}, we employ only the 12 energy bins from $i=4$ to $i=15$
of the COHERENT spectrum, because they cover the recoil kinetic energy of the more recent
Chicago-3 quenching factor measurement~\cite{Collar:2019ihs} that we use in this work. In Eq.~(\ref{chi-CsI}),
$\alpha_{\text{c}}$ and $\beta_{\text{c}}$
are nuisance parameters which quantify,
respectively,
the systematic uncertainty of the signal rate
and
the systematic uncertainty of the background rate,
with the
corresponding standard deviations
$\sigma_{\alpha_{\text{c}}} = 0.112$
and
$\sigma_{\beta_{\text{c}}} = 0.25$
\cite{Akimov:2017ade}.
The uncertainty on the quenching factor is taken into account through a normalization factor $\eta$ with $\sigma_\eta$= 0.051, that contributes to the least-squares function. 
%The value of $\sigma_{\alpha_{\text{c}}}$ is smaller than that considered in previous analyses because the previous value (0.28) included the quenching factor uncertainty, that in Eq.~(\ref{chi-CsI}) is taken into account through the factor $\eta$ in Eq.~(\ref{qfc}), with $\sigma_{\eta} = 0.051$ according to the new determination in Ref.~\cite{Collar:2019ihs}. We calculated the new value of $\sigma_{\alpha_{\text{c}}}$ by summing in quadrature the 5\% signal acceptance uncertainty and the 10\% neutron flux uncertainty estimated by the COHERENT collaboration \cite{Akimov:2017ade}, without considering an estimated 5\% neutron form factor uncertainty because we obtain the neutron form factor from the data.

Concerning the analysis of the Ar dataset, in the COHERENT paper~\cite{Akimov:2020pdx} two independent analyses are presented, labeled as A and B, that differ mainly for the selection and the treatment of the background. In the following, we will use the data coming from the analysis A, whose range of interest of the nuclear recoil energy is [0, 120]~$\mathrm{keV}_{ee}$ (corresponding to roughly [0, 350]~$\mathrm{keV}_{\mathrm{nr}}$), with 12 energy bins of size equal to 10~$\mathrm{keV}_{ee}$. 
In our analysis 
we considered the same least-squares function employed in Ref.~\cite{Cadeddu:2020lky}, namely
\begin{eqnarray}
\chi^2_{\text{Ar}}
&=&
\sum_{i=1}^{12}
\left(
\dfrac{
N_{i}^{\text{exp}}
-
\eta_{\mathrm{CE}\nu \mathrm{NS}} N_i^{\mathrm{CE}\nu \mathrm{NS}}
-
\eta_{\mathrm{PBRN}} B_i^{\mathrm{PBRN}}
-
\eta_{\mathrm{LBRN}} B_i^{\mathrm{LBRN}}}
{\sigma_i}
\right)^2\\ \nonumber
&+&
\left( \dfrac{\eta_{\mathrm{CE}\nu \mathrm{NS}}-1}{\sigma_{\mathrm{CE}\nu \mathrm{NS}}} \right)^2
+
\left( \dfrac{\eta_{\mathrm{PBRN}}-1}{\sigma_{\mathrm{PBRN}}} \right)^2
+
\left( \dfrac{\eta_{\mathrm{LBRN}}-1}{\sigma_{\mathrm{LBRN}}} \right)^2
,
\label{chi-Ar}
\end{eqnarray}
where PBRN stands for Prompt Beam-Related Background, LBRN for Late Beam-Related Neutron Background and with
\begin{eqnarray}
\sigma_i^2 = \left( \sigma_i^{\mathrm{exp}} \right)^2 &+&
\left[ \sigma_{\mathrm{BRNES}} \left( B_i^{\mathrm{PBRN}} + B_i^{\mathrm{LBRN}}\right)\right]^2,\\
\sigma_{\mathrm{BRNES}} &=& \sqrt{\frac{0.058^2}{12}}=1.7\%,\\
\sigma_{\mathrm{CE}\nu \mathrm{NS}} &=& 13.4\%\,\mathrm{for\,fixed\,}R_{n},\,\mathrm{or}\,13.2\%\,\mathrm{for\,free\,}R_{n},\\
\sigma_{\mathrm{PBRN}} &=& 32\%,\\
\sigma_{\mathrm{LBRN}} &=& 100\%.
\end{eqnarray}
For each energy bin $i$, $B_i^{\mathrm{PBRN}}$ and $B_i^{\mathrm{LBRN}}$ are the estimated number of PBRN and LBRN background events.
The Beam Related Neutron Energy Shape (BRNES) 5.8$\%$ uncertainty ($\sigma_{\mathrm{BRNES}}$) is taken into account by distributing it over the 12 bins in an uncorrelated way. 
%All the numbers are taken from Ref.~\cite{Akimov:2020pdx}. 
In Eq.~(\ref{chi-Ar}),
$\eta_{\mathrm{CE}\nu \mathrm{NS}}$, $\eta_{\mathrm{PBRN}}$ and $\eta_{\mathrm{LBRN}}$ 
are nuisance parameters which quantify,
respectively,
the systematic uncertainty of the signal rate
and
the systematic uncertainty of the PBRN and LBRN background rate,
with the
corresponding standard deviations
$\sigma_{\mathrm{CE}\nu \mathrm{NS}}$, $\sigma_{\mathrm{PBRN}}$
and
$\sigma_{\mathrm{LBRN}}$.

%In Figures 1 of Ref.~\cite{Cadeddu:2019eta,Cadeddu:2020lky} it is reported the COHERENT spectral data together with the best-fit histogram obtained with
%the SM CE$\nu$NS cross section of Eq.~(\ref{cs-std}) for CsI and Ar, respectively. A very good agreement is obtained with the SM results presented by the COHERENT Collaboration demonstrating the correct implementation of the signal and background.

\section{Constraints on light vector mediator models}
\label{sec:const}

In this section we derive the constraints that can be obtained using CsI and Ar COHERENT data on the mass, $M_{Z'}$, and the coupling, $g_{Z'}$, of a light vector mediator that couples with the SM particles according to the models described in Section~\ref{sec:light_med}.
Some of these constraints have been already derived in CsI for the universal model, see Ref.~\cite{Liao:2019uzy,Liao:2017uzy,Denton:2018xmq,Papoulias:2019xaw,papoulias2019coherent} and for the $L_\mu-L_\tau$ model~\cite{Abdullah:2018ykz}.
Here in our paper we derive the limits for CsI using the more recent Chicago-3 quenching factor measurement~\cite{Collar:2019ihs}, we derive the same limits in the recently released Ar dataset and we perform for the first time their combination.

\begin{figure*}[!t]
\centering
\setlength{\tabcolsep}{0pt}
\begin{tabular}{cc}
\subfigure[]{\label{fig:hist1}
\begin{tabular}{c}
\includegraphics*[width=0.49\linewidth]{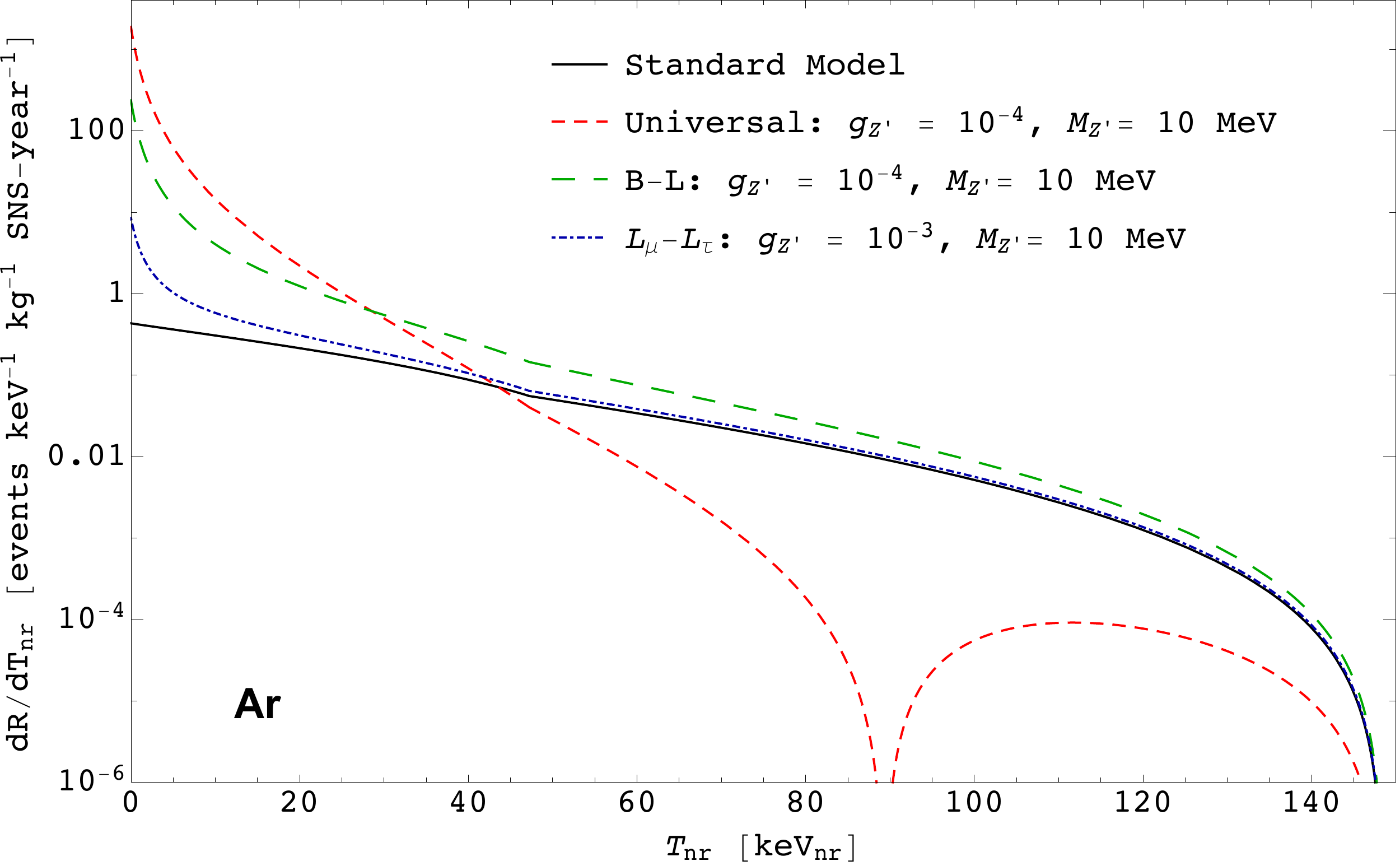}
\\
\end{tabular}
}
&
\subfigure[]{\label{fig:hist2}
\begin{tabular}{c}
\includegraphics*[width=0.49\linewidth]{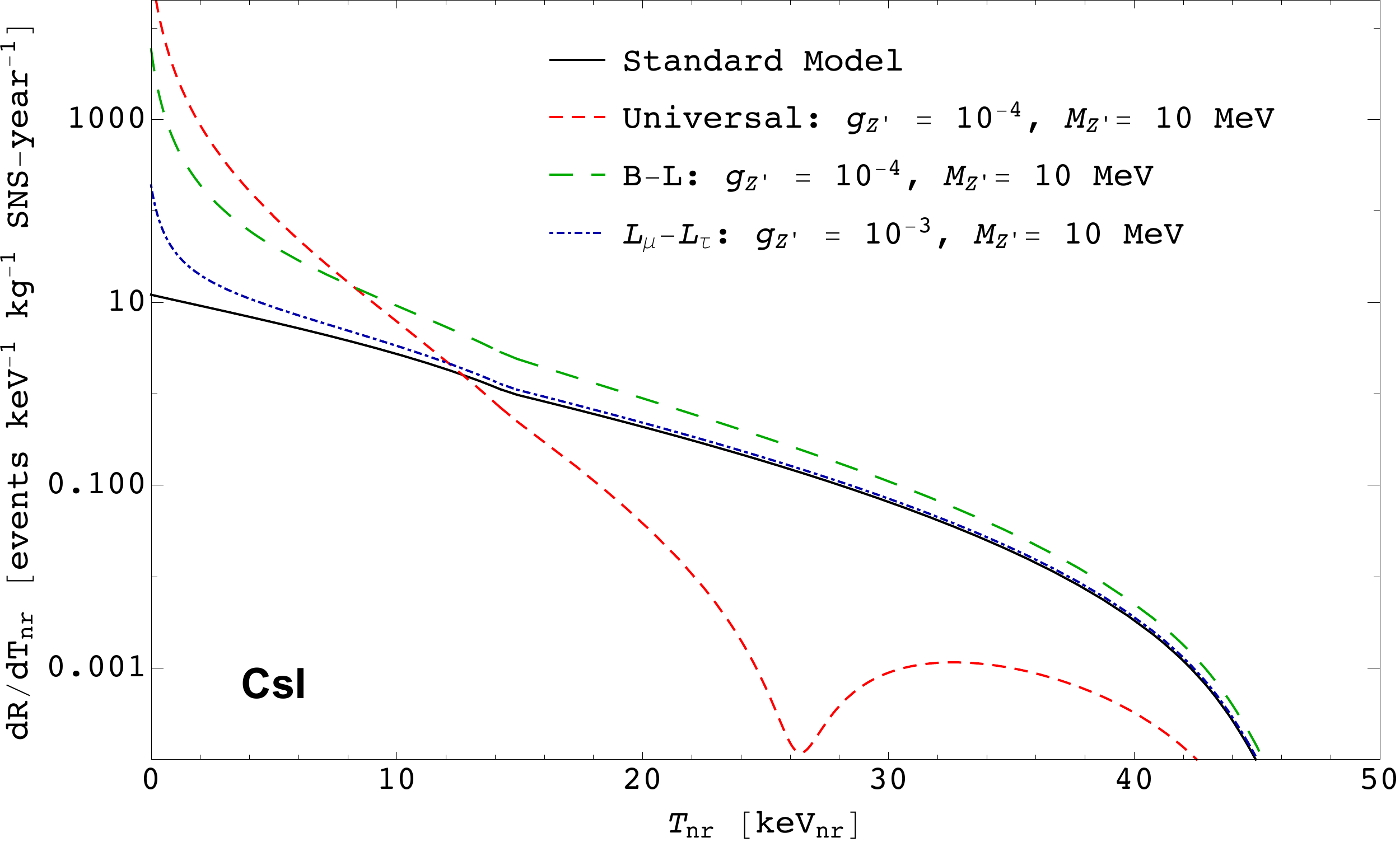}
\\
\end{tabular}
}
\end{tabular}
\caption{ \label{fig:dRdE}
Differential predicted \cenns event rates for Ar (a) and CsI (b) in the Standard Model (solid black), and universal (thin dashed red), $B-L$ (dashed green) and $L_\mu-L_\tau$ (dashed-dotted blue) models with  $M_{Z'}=10$ MeV and different couplings $g_{Z'}$ specified in the figure label.
}
\end{figure*}

First of all, in Figure~\ref{fig:dRdE} we compare the SM differential \cenns event rate for Ar (a) and CsI (b) to those obtained for the universal, $B-L$ and $L_\mu-L_\tau$ light vector mediator models.
For illustrative purposes, we fixed the mass of the light vector mediator to $M_{Z'}=10$ MeV and we use different values of the couplings $g_{Z'}$ depending on the model. The latter have been chosen close to the current lower limits determined by other experiments.
%Similarly, in Fig. \ref{fig:dRdE_binned} it is shown the binned distribution of the events for the SM and all the aforementioned models, but only for argon.
\begin{figure*}[!t]
\centering
\setlength{\tabcolsep}{0pt}
\begin{tabular}{cc}
\subfigure[]{\label{fig:Binned_Ar}
\begin{tabular}{c}
\includegraphics*[width=0.49\linewidth]{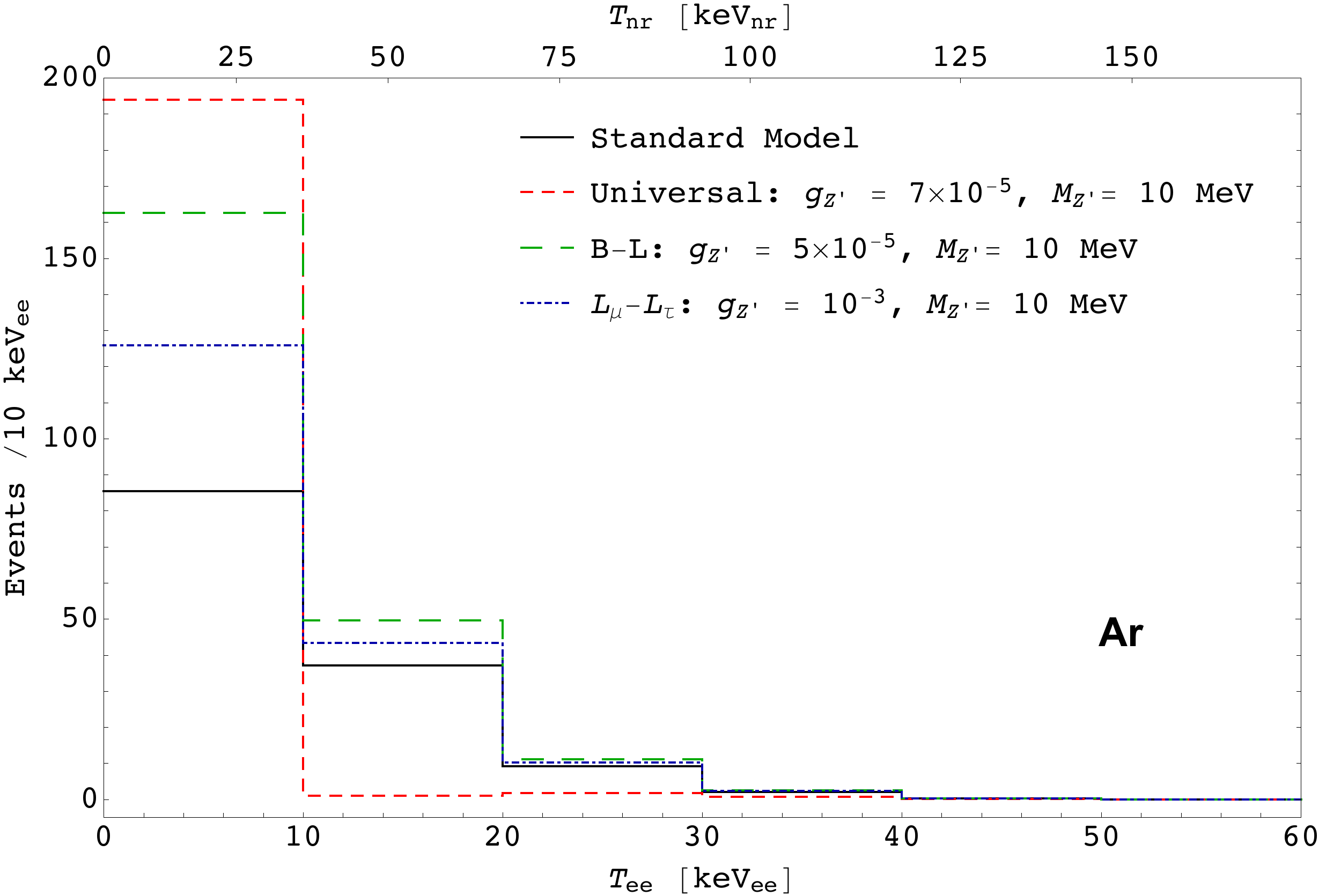}
\\
\end{tabular}
}
&
\subfigure[]{\label{fig:Binned_CsI}
\begin{tabular}{c}
\includegraphics*[width=0.49\linewidth]{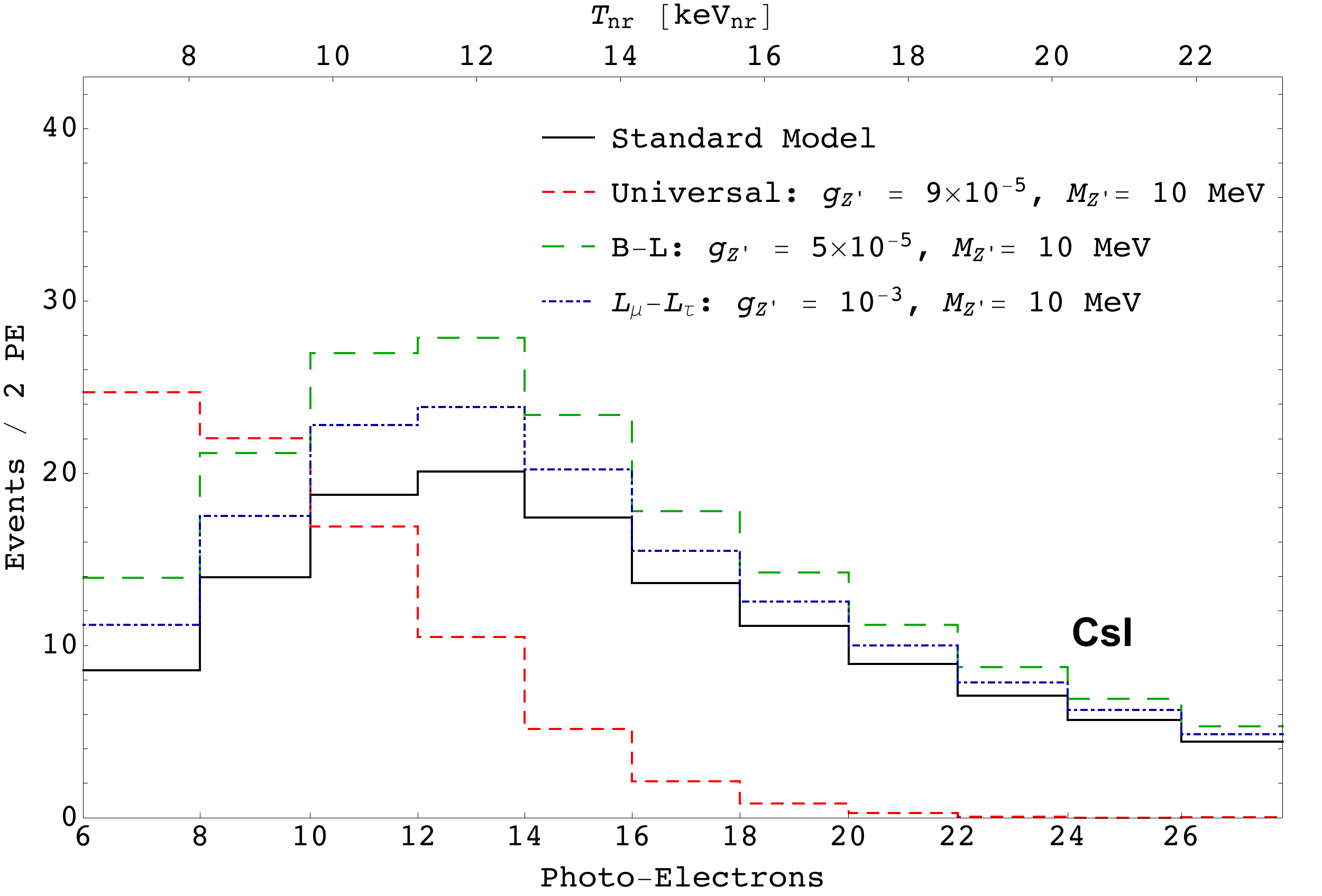}
\\
\end{tabular}
}
\end{tabular}
\caption{ \label{fig:dRdE_binned}
Binned experimental \cenns event distributions for Ar (a) and CsI (b) predicted in the Standard Model (solid black), and universal (thin dashed red), $B-L$ (dashed green) and $L_\mu-L_\tau$ (dashed-dotted blue) models with  $M_{Z'}=10$ MeV and the different values of the couplings $g_{Z'}$ specified in the figure label.
}
\end{figure*}
\begin{figure*}[t]
\centering
%\setlength{\tabcolsep}{0pt}
%\begin{tabular}{cc}
\subfigure[]{\label{fig:Coherent_Results_Universal}
%\begin{tabular}{c}
\includegraphics*[width=0.80\linewidth]{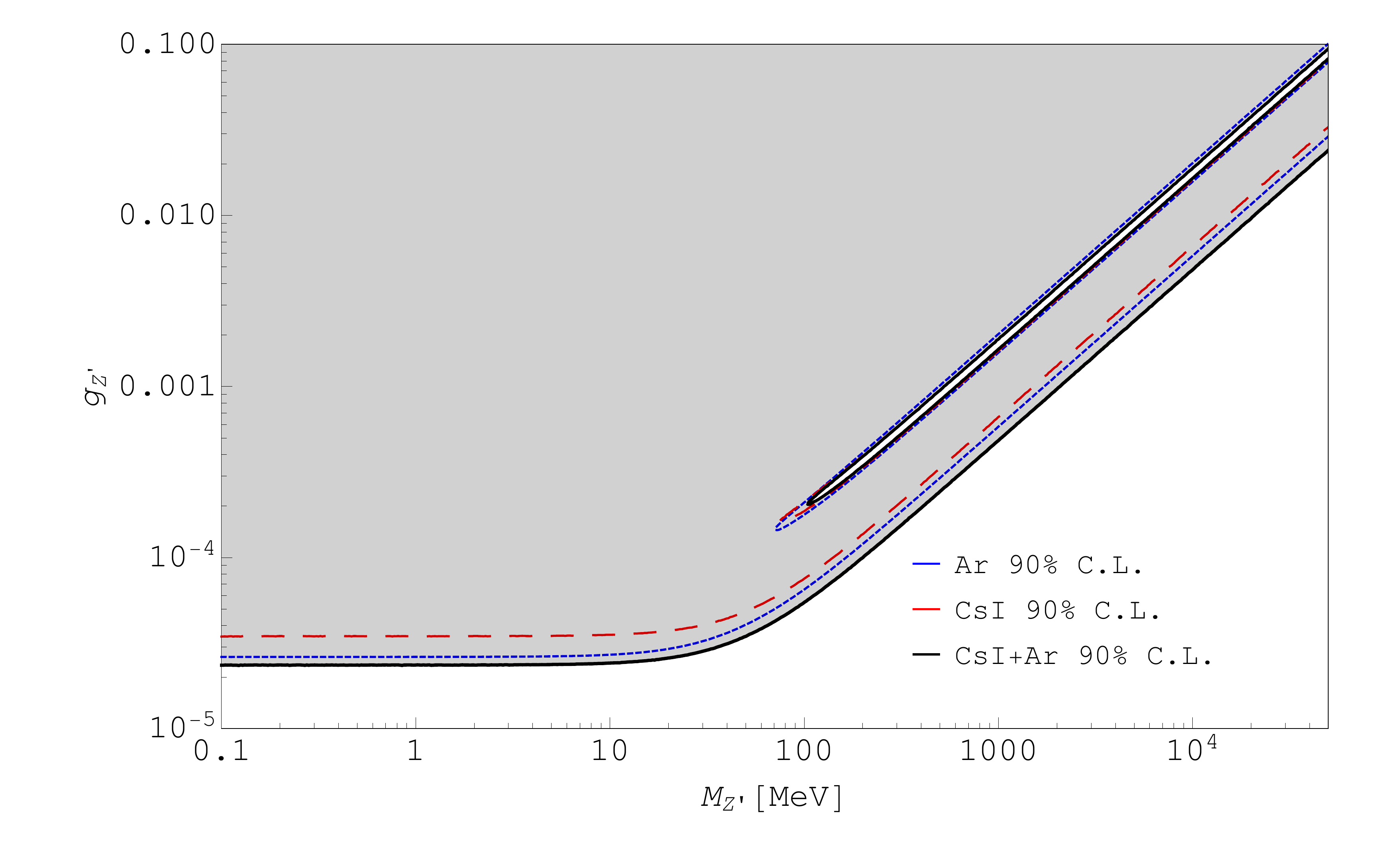}
%\\
%\end{tabular}
}
%&
\subfigure[]{\label{fig:Comparison_Universal}
%\begin{tabular}{c}
\includegraphics*[width=0.80\linewidth]{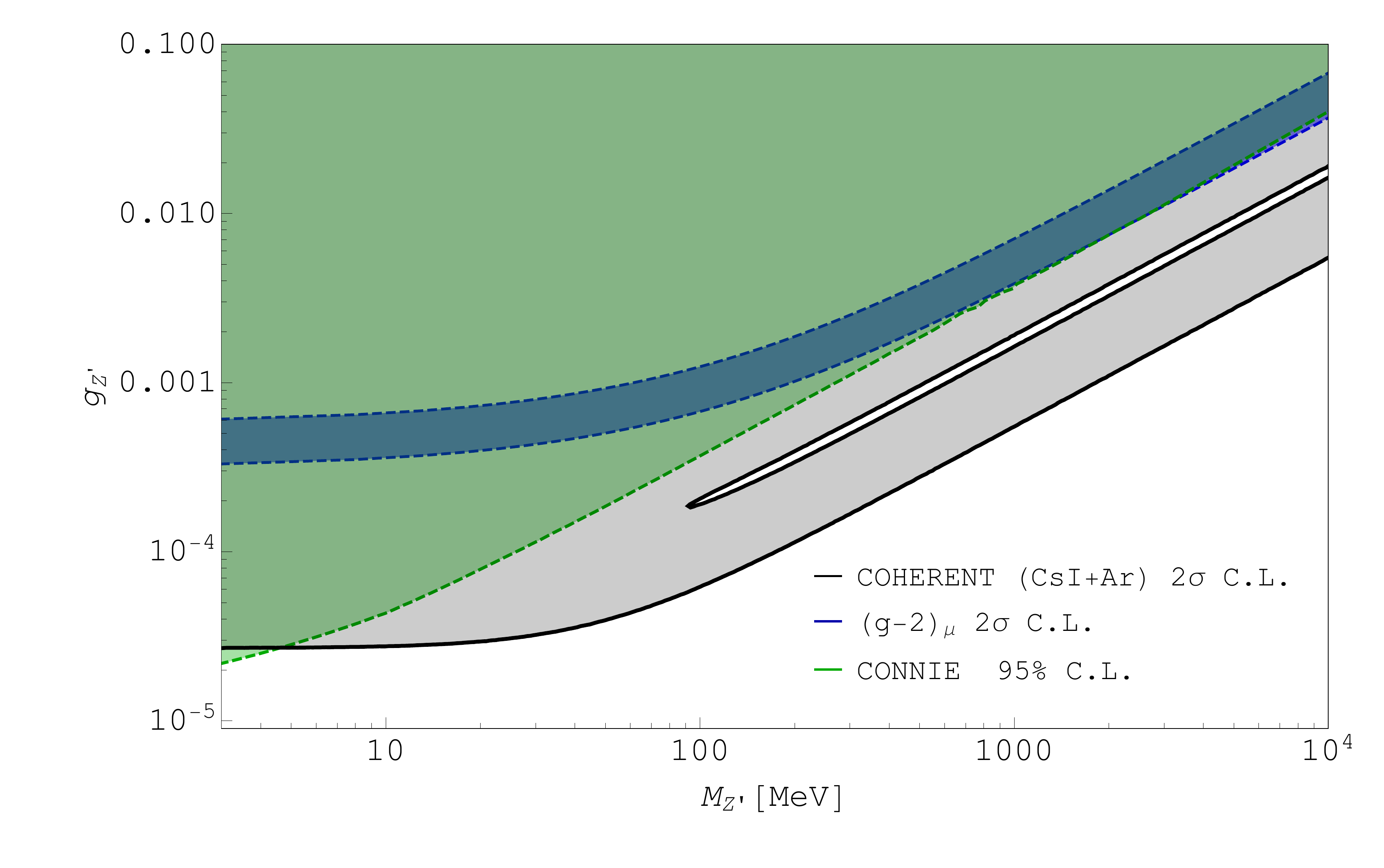}
%\\
%\end{tabular}
}
%\end{tabular}
\caption{ \label{fig:Universal_limits}
(a) Excluded regions in the $M_{Z'}-g_{Z'}$ plane for the universal model at 90\% C.L. using COHERENT Ar (limited by the dotted blue line), CsI (limited by the dashed red) data sets and their combination (solid black-grey shaded area). (b) Comparison of the combined  CsI+Ar COHERENT limits at 2$\sigma$ C.L. with the experimental bounds at 95\% C.L. from the CONNIE experiment using the Lindhard quenching factor~\cite{Aguilar_Arevalo_2020} (green area) and with the $(g-2)_{\mu}$ 2$\sigma$ region indicated by the explanation of the anomalous  magnetic moment of the muon~\cite{Bennett_2006,aoyama2020anomalous} (blue region).
}
\end{figure*}
It is possible to see that all the light vector mediator models cause an increment of the differential event rate for small recoil energies, i.e., less than about 40 $\mathrm{keV_{nr}}$ for Ar and less than about 15 $\mathrm{keV_{nr}}$ for CsI, respectively. Interestingly, the universal model also shows a dip in the event rate due to the fact that in Eq.~\ref{eq:uni} the contribution of the light vector mediator enters with opposite sign with respect to the SM one. In CsI the cancellation between the two contributions is not perfect due to the presence of two different atomic species that compensate slightly each other.

When the experimental COHERENT efficiency is taken into account for the recoil energy bins described in Section~\ref{sec:data_analysis}, the theoretical event rates transform into the experimental spectral distributions shown in Figure~\ref{fig:dRdE_binned} for Ar (a) and CsI (b). The same mass but slightly different couplings of the light vector mediator for the different models as in the previous figure have been assumed for illustrative purposes.

In Figure~\ref{fig:Coherent_Results_Universal} we report the 90\% C.L. limits obtained using Ar and CsI COHERENT data for the universal $Z'$ model, where the grey shaded area represents the region excluded by the combination of the Ar and CsI datasets. The limits obtained from the separate Ar and CsI analyses are similar, with the Ar limit being slightly more stringent for $M_{Z'} \lesssim 100$~MeV.
Note that there is a thin diagonal strip that is allowed because it corresponds to values of $g_{Z'}$ and $M_{Z'}$ for which the universal $Z'$ cross section in Eq.~(\ref{eq:uni})
is almost degenerate with the SM cross section in Eq.~(\ref{cs-std}) that fits well the data.
This happens for values of $M_{Z'}$ much larger than the typical momentum transfer
of the order of few tens of MeV.
Neglecting the $|\vec{q}|^2$ in the denominator of the $Z'$ contribution, the small proton contribution to the SM cross section,
and all the form factors,
and considering $ g_V^n \simeq - 1/2 $,
we obtain the approximate degeneracy condition
\begin{equation}
- \dfrac{N}{2}
+ \dfrac{3g_{Z'}^2}{\sqrt{2}G_F}\dfrac{Z+N}{M_{Z'}^2}
\simeq
\dfrac{N}{2}
.
\label{eq:deg1}
\end{equation}
Taking into account that
$ N / ( Z + N ) $
is about 0.55 for $^{40}\text{Ar}$
and
about 0.58 for $^{127}\text{I}$ and $^{133}\text{Cs}$,
we can write the approximate degeneracy condition as
\begin{equation}
g_{Z'} \simeq 2 \times 10^{-6} \, \dfrac{M_{Z'}}{\text{MeV}}
.
\label{eq:deg2}
\end{equation}
One can easily see that the thin diagonal allowed strip in Figure~\ref{fig:Coherent_Results_Universal}
corresponds to this approximate relation.

In Figure~\ref{fig:Comparison_Universal} the combined Ar and CsI result is shown at 2$\sigma$ C.L. to allow a better comparison with the constraints at 95\% C.L. obtained in the CONNIE reactor \cenns experiment~\cite{Aguilar_Arevalo_2020} using the Lindhard quenching factor and with the $(g-2)_{\mu}$ 2$\sigma$ region in order to explain in this model
the anomalous magnetic moment of the muon~\cite{Bennett_2006,aoyama2020anomalous}. As one can see, the CONNIE bound is the most stringent in the very low mass region of the mediator, namely for $M_{Z'}\lesssim5$~MeV. In the region above, our limits obtained from the COHERENT data improve the CONNIE limits and allow us to exclude with much higher confidence level the $(g-2)_{\mu}$ region
(the $\chi^2$ difference with the minimum is more than 120, with two degrees of freedom, that means a practical certainty of exclusion).
The high sensitivity to this model is due to the fact that the \cenns cross section is strongly modified with respect to the SM one by the introduction of the universal $Z'$ mediator with sufficiently large coupling, as visible in Figure~\ref{fig:dRdE}.

Let us also compare the limits that we obtain for CsI with those obtained in Ref.~\cite{papoulias2019coherent}, that have been derived using the same quenching factor as in this work, but considering only the total number of events in the COHERENT
CsI experiment.
Comparing our Figure~\ref{fig:Coherent_Results_Universal}
with Fig.~8 of Ref.~\cite{papoulias2019coherent},
one can see the impact of performing a spectral analysis of the COHERENT data instead of using only the total event number.
Indeed, in Fig.~8 of Ref.~\cite{papoulias2019coherent}
the degeneracy region extends down to very low $Z'$ masses,
while the spectral information allows us to restrict it to a very narrow island only present for values of $M_{Z'}$ larger than about 100~MeV.
Moreover, also the overall limit becomes more stringent by including the spectral information.
A comparable behaviour was also found in Ref.~\cite{Liao:2017uzy}, that similarly to us used the spectral information, but employed the constant quenching factor
in the COHERENT publication~\cite{Akimov:2017ade},
that has a larger uncertainty than that in Ref.~\cite{Collar:2019ihs} used by ourselves.
Comparing our Figure~\ref{fig:Comparison_Universal} to
Fig.~2 of Ref.~\cite{Liao:2017uzy}, one can see that our limits are
much more stringent
(for example, for $M_{Z'} \lesssim 10 \, \text{MeV}$,
we obtain
$g_{Z'} \gtrsim 3 \times 10^{-5}$ at $2\sigma$,
whereas the $2\sigma$ limit in Ref.~\cite{Liao:2017uzy} is
$g_{Z'} \gtrsim 6 \times 10^{-5}$).\\

\begin{figure*}%[!t]
\begin{center}
%\setlength{\tabcolsep}{0pt}
%\begin{tabular}{cc}
\subfigure[]{\label{fig:Coherent_Results_B-L}
%\begin{tabular}{c}
\includegraphics*[width=0.77\linewidth]{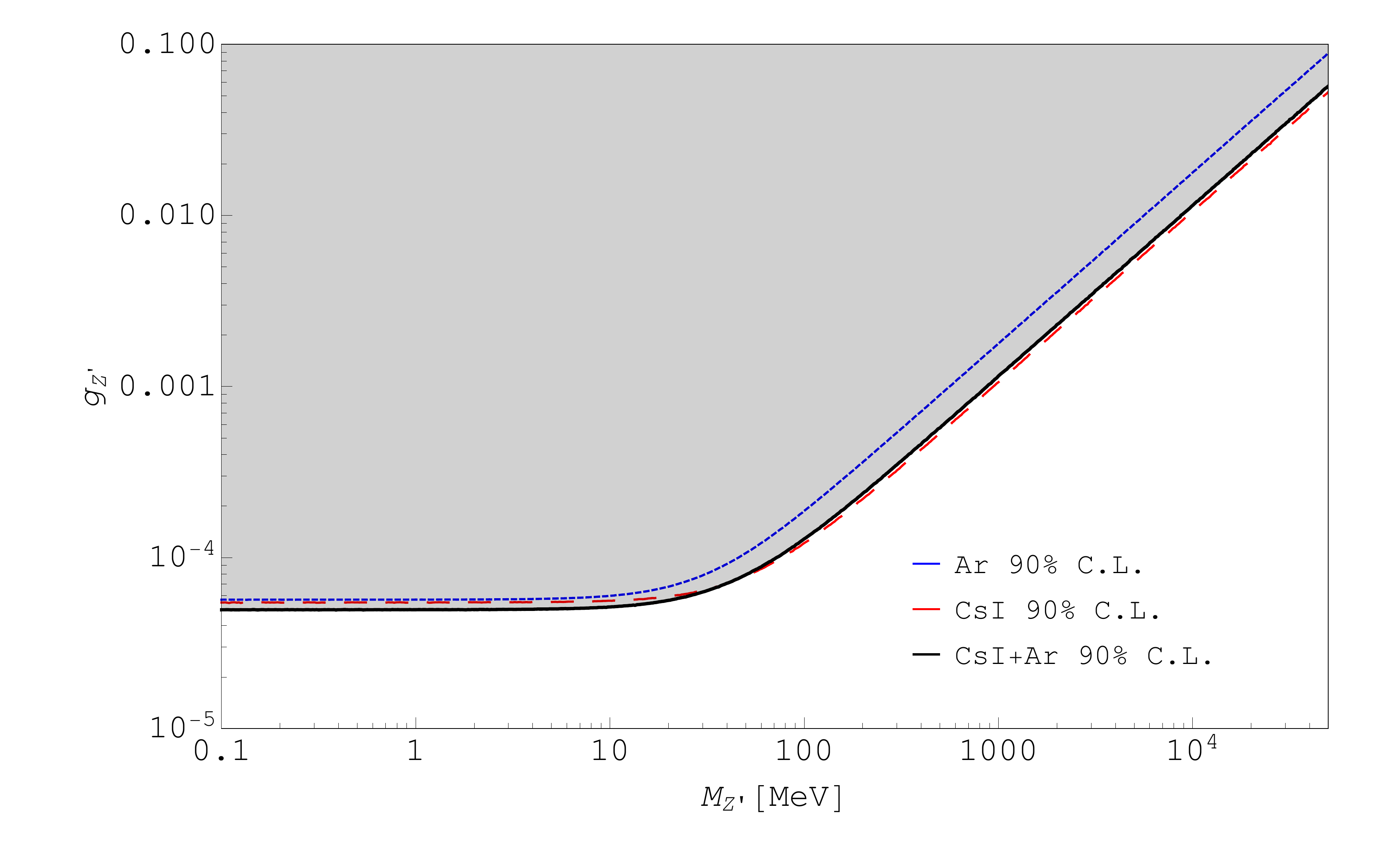}
%\end{tabular}
}\\
%&
\subfigure[]{\label{fig:Comparison_B-L}
%\begin{tabular}{c}
\includegraphics*[width=0.80\linewidth]{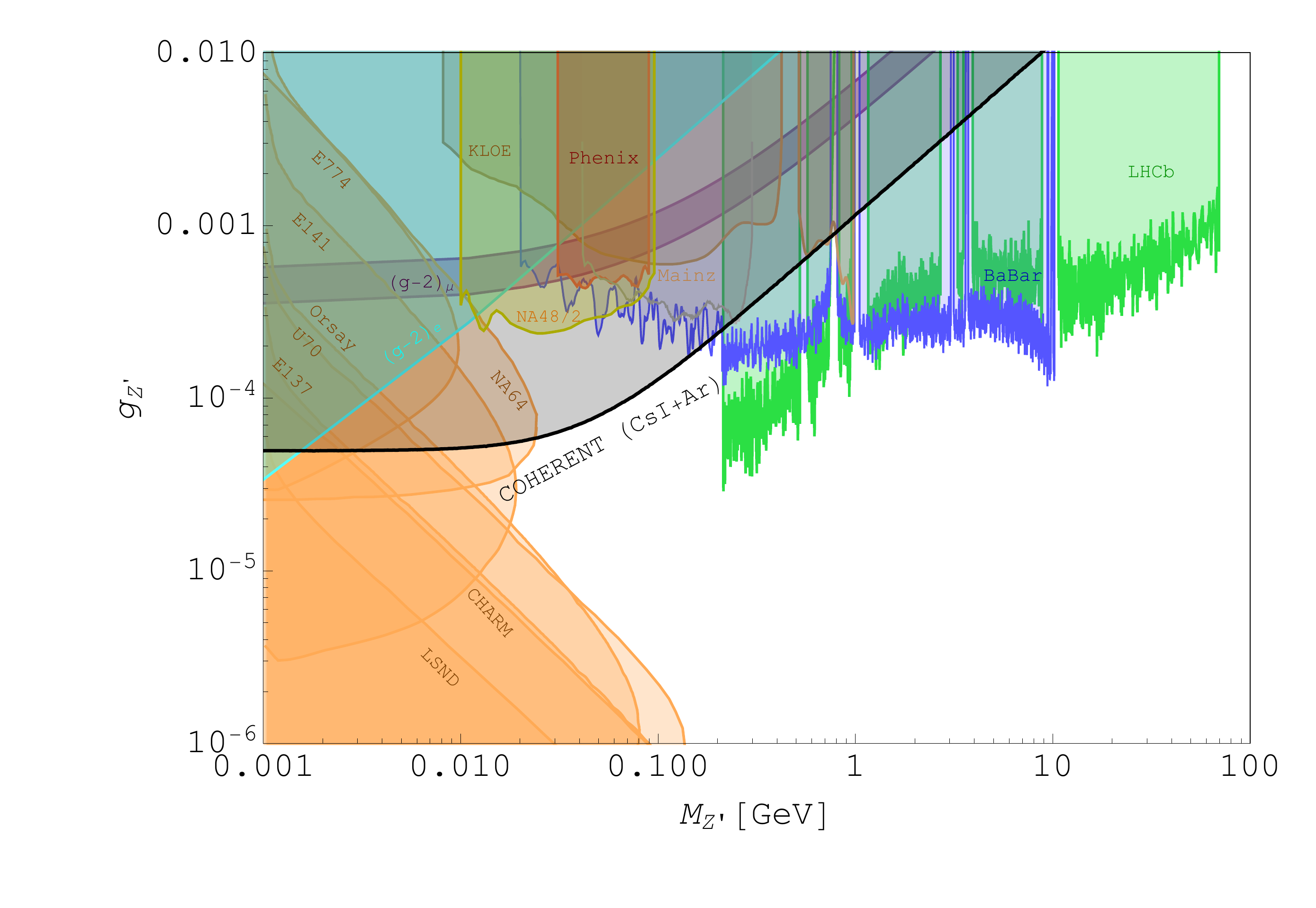}
%\\
%\end{tabular}
}
%\end{tabular}
\caption{ \label{fig:B-L_limits}
(a) Excluded regions in the $M_{Z'}-g_{Z'}$ plane for the $B-L$ model at 90\% C.L. using the COHERENT Ar (above the dotted blue line), CsI (above the dashed red line) data sets and their combination (solid black-grey shaded area). (b) Comparison of the combined COHERENT CsI+Ar 90\% C.L. excluded area with those obtained from
the BaBar~\cite{Lees_2014} (violet region), LHCb~\cite{Aaij_2019} (green region), KLOE~\cite{Abelev:2012hxa} (brown region), Phenix~\cite{PhysRevC.91.031901} (dark orange region), NA48/2~\cite{Batley:2015lha} (yellow region), A1/MAMI in Mainz~\cite{Mainz,Mainz2} (pink region) and different fixed target~\cite{Harnik_2012} (orange regions) experiments (E774~\cite{PhysRevLett.67.2942}, E141~\cite{PhysRevLett.59.755}, Orsay~\cite{DAVIER1989150}, U70~\cite{U70}, E137~\cite{PhysRevD.38.3375}, CHARM~\cite{BERGSMA1985458} and LSND~\cite{Athanassopoulos:1997er}). The $(g-2)_{\mu}$ 90\% C.L. region needed to explain the anomalous  magnetic moment of the muon~\cite{Bennett_2006,aoyama2020anomalous} (purple region) and the $(g-2)_{\mathrm{e}}$ 90\% C.L. limit~\cite{g2electron} (cyan line) from the searches of an anomalous  magnetic moment of the electron are also shown.
}
\end{center}
\end{figure*}

In Figure~\ref{fig:Coherent_Results_B-L} we show the 90\% C.L. limits obtained using the Ar and CsI COHERENT data for the $B-L$ $Z'$ model, together with their combination.
In this case, the CsI limit is more stringent than that obtained with Ar for $M_{Z'}\gtrsim10$~MeV and dominates the combined bound.
The combination of the two data sets produces only a slight improvement of the
separate limits for $M_{Z'}\lesssim10$~MeV.
In Figure~\ref{fig:Comparison_B-L} the combined CsI and Ar bound is shown together with the regions excluded at 90\% C.L. obtained by interpreting
the BaBar~\cite{Lees_2014} and LHCb~\cite{Aaij_2019}
dark photon constraints in terms of the
$B-L$ $Z'$ model~\cite{Ilten:2018crw}.
Moreover, the regions excluded at 90\% C.L. by KLOE~\cite{Abelev:2012hxa}, Phenix~\cite{PhysRevC.91.031901}, NA48/2~\cite{Batley:2015lha}, Mainz~\cite{Mainz,Mainz2} are also indicated.   
The  areas  excluded  at 90\%  C.L.  by  fixed  target  experiments~\cite{Harnik_2012} (E774~\cite{PhysRevLett.67.2942}, E141~\cite{PhysRevLett.59.755}, Orsay~\cite{DAVIER1989150}, U70~\cite{U70}, E137~\cite{PhysRevD.38.3375}, CHARM~\cite{BERGSMA1985458} and LSND~\cite{Athanassopoulos:1997er}) are also shown.
For completeness, also the $(g-2)_{\mu}$ 90\% C.L. region needed to explain the anomalous  magnetic moment of the muon~\cite{Bennett_2006,aoyama2020anomalous} and the $(g-2)_{\mathrm{e}}$ 90\% C.L. limit~\cite{g2electron} from the searches of an anomalous  magnetic moment of the electron are depicted. The COHERENT limit obtained in this work allows us to improve the coverage between accelerator and fixed target experiment limits. 

Our limits for the $B-L$ $Z'$ model are similar to those obtained in
Ref.~\cite{Miranda:2020tif},
that have been derived using only the total number of events.
In particular, it is difficult to compare our analysis of the COHERENT
Ar dataset
with that in Ref.~\cite{Miranda:2020tif}, that used the fitted number
of \cenns events derived by
the COHERENT Collaboration~\cite{Akimov:2020pdx} instead of the
experimental data.\\

\begin{figure*}%[t]
\centering
%\setlength{\tabcolsep}{0pt}
%\begin{tabular}{cc}
\subfigure[]{\label{fig:Coherent_Results_Lmu-Ltau}
%\begin{tabular}{c}
\includegraphics*[width=0.80\linewidth]{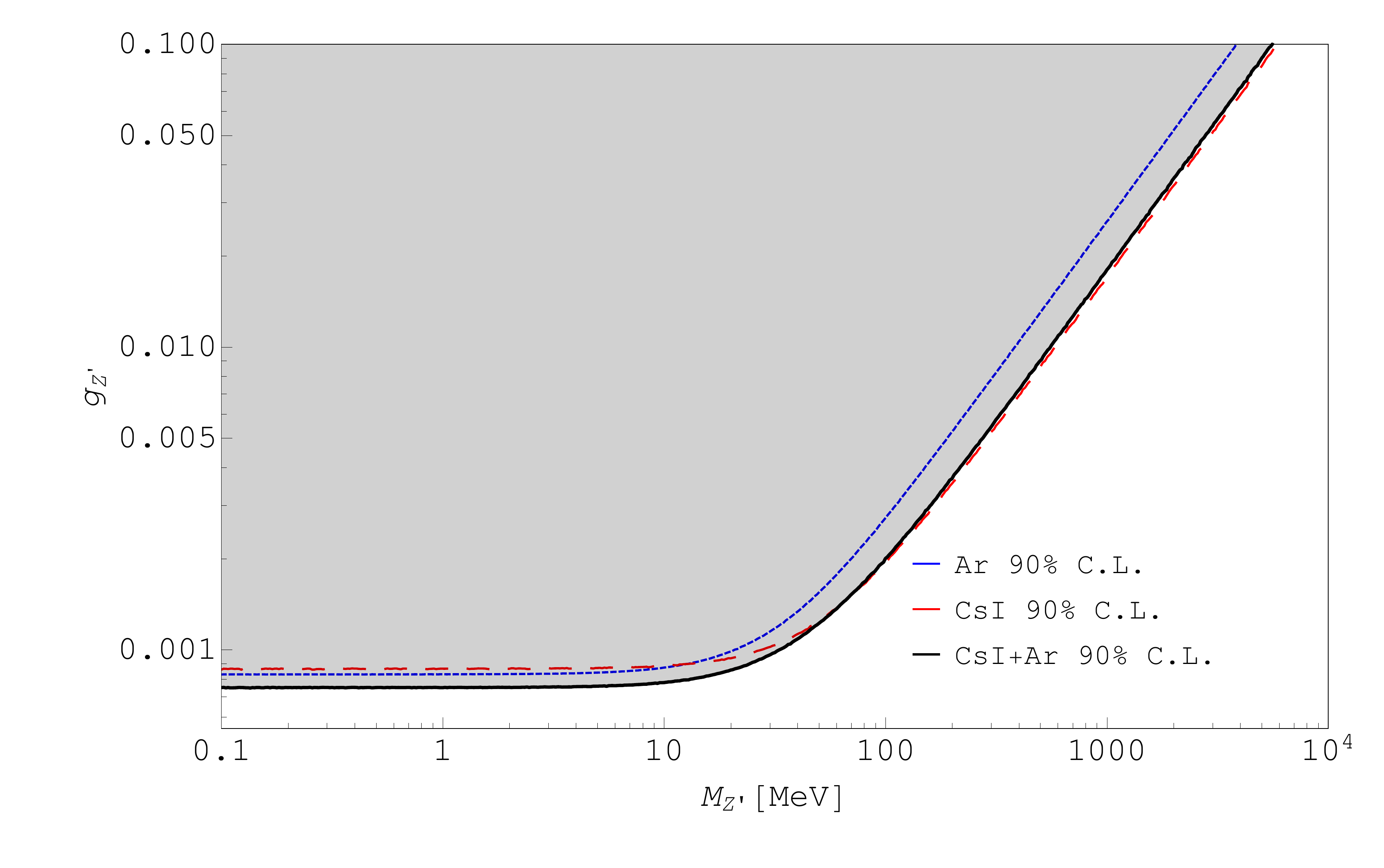}
%\end{tabular}
}\\
%&
\subfigure[]{\label{fig:Comparison_Lmu-Ltau}
%\begin{tabular}{c}
\includegraphics*[width=0.80\linewidth]{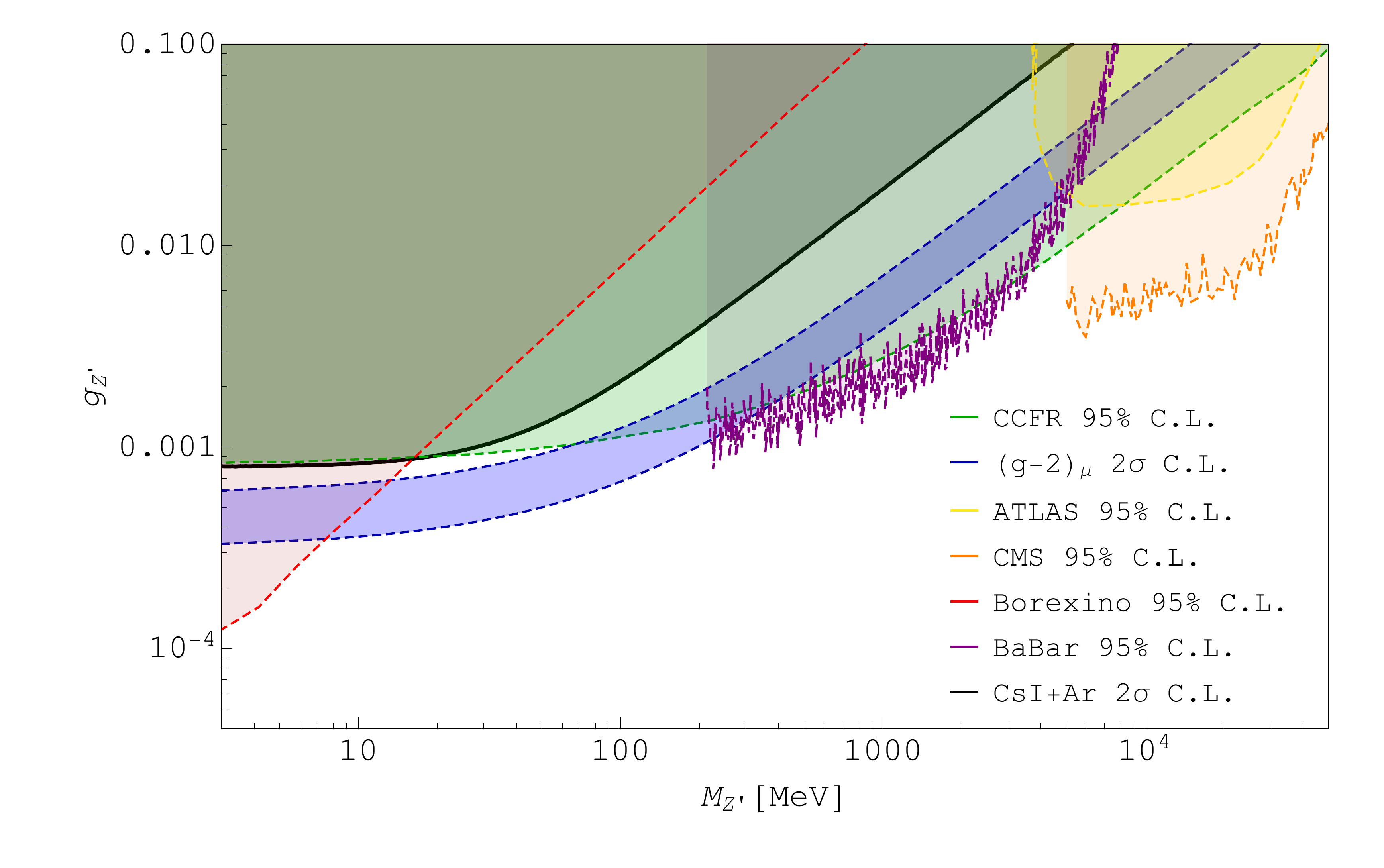}
%\\
%\end{tabular}
}
%\end{tabular}
\caption{ \label{fig:Lmu-Ltau_limits}
(a) Excluded regions in the $M_{Z'}-g_{Z'}$ plane for the $L_{\mu}-L_{\tau}$ model at 90\% C.L. using COHERENT Ar (above the dotted blue line) and CsI (above the dashed red) data sets, and their combination (solid black-grey shaded area). (b) Comparison of the combined COHERENT CsI and Ar bound at 2$\sigma$ with the existing experimental bounds at 95\% C.L. from the CCFR experiment~\cite{Altmannshofer_2014} (green area)
and from LHC searches in ATLAS~\cite{Altmannshofer:2016jzy,Aad:2014wra} (yellow area) and CMS~\cite{Sirunyan_2019} (orange area),
and the rescaled bounds at 95\% C.L. from the BaBar~\cite{TheBABAR:2016rlg} (purple area) and the Borexino~\cite{Kamada:2015era,Altmannshofer:2019zhy,Harnik_2012,Gninenko:2020xys} (red area) experiments.
The $(g-2)_{\mu}$ 2$\sigma$ region needed to explain the anomalous  magnetic moment of the muon~\cite{Bennett_2006,aoyama2020anomalous} (blue region) is also shown.
}
\end{figure*}

Finally, in Figure~\ref{fig:Coherent_Results_Lmu-Ltau} we report the 90\% C.L. limits obtained using the Ar and CsI COHERENT data for the $L_{\mu}-L_{\tau}$ $Z'$ model
and the combined bound.
In this framework, as in the $B-L$ $Z'$ case, the CsI limit is more stringent than that obtained with Ar for $M_{Z'}\gtrsim10$~MeV and dominates the combined bound.
However, the combination of the two data sets allows us to improve the separate limits for $M_{Z'}\lesssim10$~MeV slightly more than in the $B-L$ $Z'$ case.
In Figure~\ref{fig:Comparison_Lmu-Ltau} the combined CsI and Ar bound is shown together with the constraints at 95\% C.L. obtained with the CCFR measurement~\cite{Altmannshofer_2014} of the neutrino trident cross section,
with the search of SM $Z$ boson decay to four leptons in ATLAS~\cite{Altmannshofer:2016jzy,Aad:2014wra} and CMS~\cite{Sirunyan_2019} reinterpreted under the hypothesis $Z \to Z'\mu\mu$ and with the search for $e^+e^- \to \mu^+\mu^- Z',\,Z'\to \mu^+\mu^-$ from BaBar~\cite{TheBABAR:2016rlg}.
The constraints at 90\% C.L. for the Borexino~\cite{Kamada:2015era,Altmannshofer:2019zhy,Harnik_2012,Gninenko:2020xys} experiment and the $(g-2)_{\mu}$ 2$\sigma$ band, related to the  anomalous  magnetic  moment  of the muon~\cite{Bennett_2006,aoyama2020anomalous}, are also shown.
For the $L_{\mu}-L_{\tau}$ model,
the present COHERENT dataset is unfortunately unable to improve the current existing limits
and it is not sensitive to the $(g-2)_{\mu}$ band, most of which is already excluded by other data, except for $10\lesssim M_{Z'} \lesssim 400$~MeV.
However, the shape of the COHERENT combined CsI and Ar bound shows good potentiality to extend the sensitivity
in the region $10\lesssim M_{Z'} \lesssim 400$~MeV
when further data now being collected by the COHERENT experiment will be released.

Our results for the $L_{\mu}-L_{\tau}$ $Z'$ model
are more stringent than those obtained recently in Ref.~\cite{Amaral:2020tga}
by fitting the number of CE$\nu$NS events obtained by the
COHERENT Collaboration from the fit of the data~\cite{Akimov:2020pdx}.
Since this is not a real fit of the COHERENT data,
the result is questionable.

%Borexino B-L 

\section{Conclusions}
\label{sec:conclusions}

In this paper we discussed the limits on different
light vector mediator models
that can be obtained from the analysis of the recent
CE$\nu$NS data on argon~\cite{Akimov:2020pdx} and cesium iodide~\cite{Akimov:2017ade} of the COHERENT experiment. We also presented the results obtained by combining the analysis
of the CsI and Ar data sets.
We considered three models with a light $\mathrm{Z'}$ mediator:
one in which the $\mathrm{Z'}$ couples universally to all SM fermions,
another corresponding to a $B-L$ extension of the SM,
and a third one with a gauged $L_\mu-L_\tau$ symmetry.

We compared the results obtained from the combination of the COHERENT
CsI and Ar data sets with the limits derived from searches in fixed target, accelerator, solar neutrino and reactor \cenns experiments.
We showed that for the universal and the $B-L$ $\mathrm{Z'}$ models, the COHERENT data allow us to put stringent limits on the light vector mediator mass, $M_{Z'}$, and coupling, $g_{Z'}$, parameter space.
In particular,
in the framework of the universal $Z'$ model
we improved significantly the limits obtained in the CONNIE reactor \cenns experiment~\cite{Aguilar_Arevalo_2020}
and
we excluded, with very high confidence level, the possibility of explaining the anomalous magnetic moment of the muon~\cite{Bennett_2006,aoyama2020anomalous} assuming this model.
Considering the $B-L$ $\mathrm{Z'}$ model,
we have shown that the COHERENT data allow us to improve the coverage between accelerator and fixed target experiment limits. 
Finally, our results for the $L_{\mu}-L_{\tau}$ $Z'$ model
do not improve the  current existing limits,
but show good potentiality of future COHERENT data
to extend the sensitivity
in the region $10\lesssim M_{Z'} \lesssim 400$~MeV
and $g_{Z'} \lesssim 8 \times 10^{-4}$
where the explanation with this model of the anomalous magnetic moment of the muon is still unconstrained.

\begin{acknowledgments}
The work of C. Giunti was partially supported by the research grant "The Dark Universe: A Synergic Multimessenger Approach" number 2017X7X85K under the program PRIN 2017 funded by the Ministero dell'Istruzione, Universit\`a e della Ricerca (MIUR).
The work of Y.F. Li and Y.Y. Zhang is supported by the National Natural Science Foundation of China under Grant No.~11835013 and by the Beijing Natural Science Foundation under Grant No.~1192019. Y.F. Li is also grateful for the support by the CAS Center for Excellence in Particle Physics (CCEPP).
\end{acknowledgments}

\bibliographystyle{JHEP}
\bibliography{main}

\newpage
%\appendix
%\section{Results obtained with the official coherent quenching factor}
%\label{app:qfcoherent}

\end{document}